# Calibrationless Reconstruction of Uniformly-Undersampled Multi-Channel MR Data with Deep Learning Estimated ESPIRiT Maps

Junhao Zhang[1,2], Zheyuan Yi[1,2,3], Yujiao Zhao[1,2], Linfang Xiao[1,2], Jiahao Hu[1,2,3], Christopher Man[1,2], Vick Lau[1,2], Shi Su[1,2], Fei Chen[3], Alex T.L.Leong[1,2], and Ed X. Wu[1,2*]

[1]Laboratory of Biomedical Imaging and Signal Processing
The University of Hong Kong, Hong Kong SAR, People's Republic of China
[2]Department of Electrical and Electronic Engineering
The University of Hong Kong, Hong Kong SAR, People's Republic of China
[3]Department of Electrical and Electronic Engineering
Southern University of Science and Technology, Shenzhen, People's Republic of China

[*]Correspondence to:
        Ed X. Wu, Ph.D.
        Department of Electrical and Electronic Engineering
        The University of Hong Kong, Hong Kong SAR, China
        Tel: (852) 3917-7096
        Email: ewu@eee.hku.hk

# ABSTRACT

**Purpose:** To develop a truly calibrationless reconstruction method that derives ESPIRiT maps from uniformly-undersampled multi-channel MR data by deep learning.

**Methods:** ESPIRiT, one commonly used parallel imaging reconstruction technique, forms the images from undersampled MR k-space data using ESPIRiT maps that effectively represents coil sensitivity information. Accurate ESPIRiT map estimation requires quality coil sensitivity calibration or autocalibration data. We present a U-Net based deep learning model to estimate the multi-channel ESPIRiT maps directly from uniformly-undersampled multi-channel multi-slice MR data. The model is trained using fully-sampled multi-slice axial brain datasets from the same MR receiving coil system. To utilize subject-coil geometric parameters available for each dataset, the training imposes a hybrid loss on ESPIRiT maps at the original locations as well as their corresponding locations within the standard reference multi-slice axial stack. The performance of the approach was evaluated using publicly available T1-weighed brain and cardiac data.

**Results:** The proposed model robustly predicted multi-channel ESPIRiT maps from uniformly-undersampled k-space data. They were highly comparable to the reference ESPIRiT maps directly computed from 24 consecutive central k-space lines. Further, they led to excellent ESPIRiT reconstruction performance even at high acceleration, exhibiting a similar level of errors and artifacts to that by using reference ESPIRiT maps.

**Conclusion:** A new deep learning approach is developed to estimate ESPIRiT maps directly from uniformly-undersampled MR data. It presents a general strategy for calibrationless parallel imaging reconstruction through learning from coil and protocol specific data.

# INTRODUCTION

Magnetic Resonance Imaging (MRI) is a critical technology in modern medicine. MRI can generate anatomical and functional images with high resolution and flexible contrasts[1,2]. However, MRI comes at the expense of long scan time when compared to other medical imaging modalities. To speed up MRI data acquisition, parallel imaging has been developed and used routinely nowadays in the clinic[3]. It takes advantage of the spatial encoding effect of the MR multi-channel receiving coil system. Two seminal parallel imaging reconstruction methods are sensitivity encoding (SENSE) in image space[4] and generalized partially parallel acquisitions (GRAPPA) in k-space[5]. SENSE uses prior knowledge of coil sensitivity profiles to separate the folded pixels in the image space that arise from the uniform undersampling in k-space. GRAPPA synthesizes the missing k-space data using the GRAPPA weight kernels across all channels. Both approaches need calibration data either from additional pre-scan or autocalibration signals. Often there exists inconsistency between calibration data and undersampled data, e.g., due to subject motion, causing artifacts in the reconstructed images[6,7]. ESPIRiT[8], following SENSE and GRAPPA, is an effective hybrid-space reconstruction method. It utilizes k-space kernel operations to derive a set of eigenvector maps, i.e., ESPIRiT maps, to effectively represent coil sensitivity information. They are then incorporated in a generalized SENSE reconstruction. ESPIRiT reconstruction with L1 regularization can lead to improved performance when compared to the traditional SENSE and GRAPPA[8]. However, the ESPIRiT reconstruction still requires quality coil calibration or autocalibration data, hindering its robust applications in various imaging scenario. For instance, in abdominal[9] and cardiac parallel imaging[10,11], the geometric mismatch could easily occur due to motion between pre-scan coil sensitivity calibration data and subsequent imaging data, degrading reconstruction quality. Additionally, fast-spin-echo (FSE) and echo-planar-imaging (EPI) are among the most common template sequences in modern MR scanners due to their high acquisition efficiency. Integration of autocalibration in FSE sequence can complicate the phase encoding ordering and requires careful optimization[12,13]. For EPI based parallel imaging, it is also difficult to integrate autocalibration in the sequence because uniform undersampling in phase encoding direction is often a pre-requisite for effective reduction of geometric distortions due to field inhomogeneity[14,15].

Several calibrationless MRI reconstruction methods have been proposed for parallel imaging[16-20]. For example, low-rank reconstruction methods have emerged as the alternatives, enabling simultaneous autocalibration and k-space approximation. They include simultaneous auto-calibrating and k-space estimation (SAKE), parallel-imaging low-rank matrix modeling of local k-space neighbors (P-LORAKS) and annihilating filter-based low-rank Hankel matrix approach (ALOHA)[21-23]. These methods construct the entire undersampled k-space data into a structured low-rank matrix to recover missing k-space data through an iterative procedure using k-space data consistency and rank truncation. They exploit the inherent nature of multi-channel data and finite image spatial support. Although powerful, many such low-rank completion methods are computationally demanding, hampering their applications for high-resolution volume imaging or direct 3D reconstruction[24,25]. Note that the low-rank based ENLIVE method[26] combines ESPIRiT and NLINV[27]. It is computationally efficient, yet still requires variable density sampling and cannot accomondate uniform undersampling.

Deep learning has demonstrated tremendous success in various fields and shown great potential to significantly advance MRI[28,29]. For example, GRAPPA-Net is a full end-to-end convolutional neural network (CNN) model that first fills the missing k-space lines using non-linear CNN-based interpolation functions and then maps the filled k-space to the corresponding image space[30]. Variational network are proposed to learn the non-linear mapping from aliased image to alias-free image, where the deep learning model is embedded in the iterative image reconstruction with generalized compressed sensing concept[31]. RAKI, another end-to-end reconstruction model, reconstructs the fully-sampled MR k-space from undersampled MR k-space data[32]. In general, these parallel imaging reconstruction methods explicitly require the additional consecutive central k-space lines or certain level of coil sensitivity information. More recently, DeepSENSE[33] is proposed to reconstruct undersampled k-space data through deep learning of coil sensitivity maps, representing a new deep learning strategy in coil sensitivity map space. Yet it is not entirely calibrationless since it aims to estimate high-resolution coil sensitivity maps from the low-resolution ones that are derived from a small number of consecutive central space lines. Given that simple uniform undersampling in k-space can be and has been widely integrated into various clinical imaging protocols for acceleration or artifact correction, it is imperative to develop deep learning methods for predicting coil sensitivity information directly from uniformly-undersampled k-space data without any calibration data.

In this study, we propose a calibrationless reconstruction framework that directly estimates ESPIRiT maps from multi-channel uniformly-undersampled MR data via deep learning and applies such maps for subsequent ESPIRiT reconstruction. For demonstration, a U-Net based model for axial reconstruction is trained using fully-sampled multi-slice axial brain datasets from the same MR receiving coil system. To utilize subject-coil geometric parameters available for each dataset, the training imposes a hybrid loss on ESPIRiT maps at the original locations and their corresponding or transformed locations within the standard reference multi-slice axial stack (i.e., with a fixed geometric relation to the coil system). The results show that the deep learning model could reliably estimate the ESPIRiT maps from uniformly-undersampled axial multi-slice MR data, led to excellent ESPIRiT image reconstruction performance.

# METHODS

## Proposed deep learning estimated ESPIRiT maps for ESPIRiT reconstruction

ESPIRiT takes advantage of the autocalibration feature of GRAPPA to derive SENSE-like relative maps related to coil sensitivity, i.e., ESPIRiT maps. These maps are mathematically derived from the null space of the calibration matrix from the coil calibration data by singular value decomposition (SVD)[8,34].

The framework of our proposed ESPIRiT reconstruction is shown in **Figure 1A**. It mainly consists of two steps: estimating first set of multi-channel ESPIRiT maps from uniformly-undersampled MR data and applying the estimated maps for ESPIRiT reconstruction. Specifically, a deep learning model is developed for mapping aliased images to the corresponding ESPIRiT maps (**Figure 1B**). The input of the model is the multi-channel aliased MR images from uniformly-undersampled multi-channel data, while the output is the corresponding multi-channel ESPIRiT maps.

An attention U-Net model[35-37] is adopted (**Figure 1C**). It constitutes a key element of our proposed architecture. Note that the complex-valued input and target/output are treated as separate two real-valued channels for real and imaginary parts, respectively[31,38,39]. The number of filter channels in U-Net model is 12, 64, 128, 256, 512 and 1024 for the convolutional layers from undersampled data to latent feature spaces while 1024, 512, 256, 128, 64 and 12 for layers from latent features to multi-channel ESPIRiT maps. Additionally, channel-wise attention block is included in the model to help effectively process the input information across multiple channels[40].

The coil sensitivity information in any MRI system is largely coil-specific. When scanning a particular subject in clinical MRI setting, the exact coil sensitivity profiles or ESPIRiT maps within any imaging slice also depend on the orientation/position of the slice with respect to MR receiving coil system. For example, typical multi-slice axial head scan often involves slightly different orientation/position with respect to the standard reference multi-slice axial stack geometry (i.e., the MR coil system, too) due to variation in head position and slice localization by scanner operator (**Figure 1D**). Such subject-coil geometry information is available during the scan and recorded in the standard DICOM header. During the training of our proposed model, we incorporate these subject-coil parameters for each dataset by imposing a hybrid loss function. Specifically, the model is trained by minimizing a hybrid L1 loss on two sets of multi-slice multi-channel ESPIRiT maps as described in Equation (1).

$$\arg\min_{\theta} \sum_{ij} \left[ \lambda \left| E_{DL}^{ij} - E_{original}^{ij} \right| + (1-\lambda) \left| E_{DL}^{ij} - E_{transformed}^{ij} \right| \right] \qquad (1)$$

ZHere $\theta$ are all the parameter sets in the model and $\lambda$ is a learnable parameter to control the loss contributions. Specifically, $E^{ij}_{original}$ and $E^{ij}_{transformed}$ represent two ESPIRiT maps for $i^{th}$ channel at $j^{th}$ slice within their original multi-slice locations and their transformed locations within the standard reference multi-slice axial stack, corresponding to blue and red stacks in **Figure 1D**, respectively. Note that this reference stack has a fixed orientation and position relative to magnet and gradient coil center. Its orientation and position typically have a fixed geometric relation to the coil system. Thus $E^{ij}_{transformed}$ should be mostly coil specific and dataset independent. Meanwhile $E^{ij}_{original}$ will be dataset dependent since each multi-slice axial head scan can be prescribed with slightly different geometry. In practice, $E^{ij}_{transformed}$ and $E^{ij}_{original}$ differ from each other in position and orientation but will be very similar to certain extent due to their geometric proximity and the spatial smoothness nature of ESPIRiT maps. Note that the second term in Equation (1) here could be considered as an indirect way to initialize the model training. Therefore, incorporating $E^{ij}_{transformed}$ as part of the loss function for training will facilitate the learning process through improving stabilization and convergency indirectly. With such hybrid loss, the model can effectively learn to predict ESPIRiT maps, $E^{ij}_{DL}$, from each undersampled multi-slice dataset itself despite its varying subject-coil geometry during scan.

## Data preparation

The brain data used in our deep learning experiment are from the publicly available Calgary-Campinas MR database (https://sites.google.com/view/calgary-campinas-dataset)[41]. They included 65 fully-sampled human brain datasets from 65 individual healthy subjects that were acquired on a 1.5T clinical scanner (GE Healthcare, Waukesha, WI) using a 12-channel head coil. The T1-weighed (T1W) datasets were acquired using a 3D gradient-recalled echo (GRE) sequence with TE/TR/TI = 6.3/2.6/650 ms or TE/TR/TI = 7.4/3.1/400 ms, FOV = 256x218x180 mm$^3$ and matrix size = 128x128x80. Each dataset consisted of consecutive multi-slice multi-channel 2D data. All datasets were compressed to 6 channels for fast implementation in this study using the coil combination procedure[42].

65 datasets were divided into 50, 5 and 10 sets for training, validation, and testing respectively. Each dataset had 80 slices, providing the whole brain coverage. The stacked multi-slice multi-channel 2D k-space data were undersampled retrospectively with uniform undersampling patterns at different acceleration factors (i.e., R = 2, 3 and 4), from which aliased multi-slice multi-channel 2D images were generated as model input through zero padding of missing k-space data and simple 2D FT. For training, multi-slice multi-channel $E^{ij}_{original}$ maps were computed directly from the 24 consecutive central k-space lines that were extracted from the original fully-sampled k-space data, with kernel size 6x6 using SPIRiT software package V0.3 [8,21,42,43]. $E^{ij}_{transformed}$ maps were obtained by first transforming the original alias-free multi-slice image data to the standard reference multi-slice axial stack, and then finding the ESPIRiT maps at the corresponding or transformed locations in the new coordinate (see corresponding $E^{ij}_{original}$ and $E^{ij}_{transformed}$ locations in **Figure 1D**). The coordinate transformation was performed using the rigid-body rotation and translation described in Equation (2).

$$\begin{vmatrix} x' \\ y' \\ z' \end{vmatrix} = \begin{vmatrix} 1 & 0 & 0 \\ 0 & \cos\alpha & -\sin\alpha \\ 0 & \sin\alpha & \cos\alpha \end{vmatrix} \begin{vmatrix} \cos\beta & 0 & \sin\beta \\ 0 & 1 & 0 \\ -\sin\beta & 0 & \cos\beta \end{vmatrix} \begin{vmatrix} \cos\gamma & -\sin\gamma & 0 \\ \sin\gamma & \cos\gamma & 0 \\ 0 & 0 & 1 \end{vmatrix} \begin{vmatrix} x \\ y \\ z \end{vmatrix} + \begin{vmatrix} m \\ n \\ t \end{vmatrix} \quad (2)$$

Parameters $\alpha$，$\beta$ and $\gamma$ denote three rotation angles (pitch, roll, head rotation) and variables m, n and t represent three translation parameters (left-right, anterior-posterior, and head-foot translation) for alignment to the standard reference multi-slice axial stack. These subject-coil geometry parameters were obtained from the DICOM tags (0020,0037) and (0020,0032), respectively, of each dataset. These parameters are summarized for all 65 subjects in **Supporting Information Figure S1A.**

To further evaluate the robustness of the proposed method, cardiac MRI data from OCMR public database[44] were also used to train a model for ESPIRiT reconstruction of cardiac images. They were 28-channel multi-slice multi-phase fully-sampled 1.5 T cardiac data from 8 subjects acquired using SSFP sequence with matrix size 320x120, slice number 12, 10 cardiac phases and TR/TE = 28.5/1.4ms. Additional cropping of the slice to 120 x 120 was performed to remove some background regions. Coil compression was used to compress the data to 6 channels. Reference ESPIRiT maps were computed from 25 consecutive central k-space lines.

### Model training, testing and performance evaluation

The training process was achieved by minimizing a hybrid L1 loss in Equation (1). Adam optimizer[45] was carried out for training with $\beta_1$ = 0.9, $\beta_2$ = 0.999 and initial learning rate = 0.0001. $\lambda$ was initialized to 0.5 and gradually decreased to zero during training. As shown in **Supporting Information Figure S1B**, the weight of the second term in Equation (1), i.e., 1 - $\lambda$, decreased rapidly during the training. This indicates that, with each epoch or iteration, the model training focuses more the first term in Equation (1) instead of the second term. The training was conducted on a Geforce RTX 3090 GPU using PyTorch 1.8.1 package[46] with a batch size of 32 and 100 epochs. The total training time was approximately 27.8 hrs and 7.4 hrs, respectively for the brain T1-weighted multi-slice 6-channel model and the cardiac multi-slice 6-channel model. Note that, all cardiac data of different phases were used to the cardiac model.

For final image reconstruction, the estimated ESPIRiT maps were directly used in L1-ESPIRiT reconstruction using the SPIRiT software package V0.3[7,13,27]. For evaluation, we compared the deep learning estimated 2D ESPIRiT maps with the reference maps derived from 24 consecutive central k-space lines. We quantified the pixel-wise Pearson correlation similarity[47] between the deep learning estimated and reference maps. To assess the image reconstruction performance using the estimated maps, residual error images, normalized root-mean-square error (NRMSE)[48] and peak signal-to-noise ratio (PSNR)[49] were calculated to quantify the differences between the images reconstructed using ESPIRiT reconstruction and reference images reconstructed from fully-sampled k-space data. The image local error maps[8] were also computed.

Our proposed method and evaluation above were implemented using MatLab (MathWorks, Natick, MA). All source code can be obtained online (https://github.com/bjzhang/DeepLearningESPIRiT) or from the authors upon request.

# RESULTS

The proposed model robustly predicted multi-channel ESPIRiT maps and led to successful image reconstruction from uniformly-undersampled k-space data for all test subjects without any coil calibration data. **Figure 2** shows the typical results of deep learning estimated ESPIRiT maps at three acceleration factors (i.e., R = 2, 3 and 4) in one slice location from one subject. Visually, they were highly comparable in both magnitude and phase to the reference ESPIRiT maps derived from 24 consecutive central lines in the fully-sampled k-space (**Figure 2A**). Quantitatively, there existed a high degree of pixel-wise correlation between the deep learning estimated and reference ESPIRiT magnitude maps (**Figure 2B**). The correlation coefficients ranged from 0.96 to 0.98 at different acceleration factors across all 6 channels. With increasing R, no significant reduction in correlation coefficients was observed across 6 channels. **Figure 2C** presents the correlation coefficients between estimated and reference ESPIRiT magnitude maps across all channels from one slice at similar locations among all 10 test subjects. They were consistently high, indicating the robustness of the proposed deep learning estimation of ESPIRiT maps.

**Figure 3** demonstrates the axial brain image reconstruction performance using estimated 6-channel ESPIRiT maps by comparing with reconstruction using reference ESPIRiT maps for the same subject shown in **Figure 2** at R = 2, 3 and 4. Residual error images and their NRMSE/PSNR values are also shown. Even at high acceleration R = 4, the reconstruction using reference and estimated ESPIRiT maps yielded highly comparable performance in terms of PSNR and NRMSE. The corresponding image local error maps were also computed (**Supporting Information Figure S4**), showing artifact patterns consistent to these in the image error maps in Figure 3. In general, similar noise amplification and artifact levels were observed. Note that, at high acceleration R = 4, the reconstruction images using reference ESPIRiT maps produced slightly less aliasing than that using estimated reference maps, but showed slightly more noise amplification.

We examined the performance of our proposed method for axial brain datasets acquired with large subject-coil geometry deviation from the standard reference multi-slice axial stack orientation/position. **Figure 4** shows the reconstructed images using reference and estimated ESPIRiT maps from one subject with a large pitch rotation angle (approximately -10 degrees) at R = 4. Four slices covering a large portion of the brain are shown. The reconstructed images at each slice location using reference and deep learning estimated ESPIRiT maps were again highly comparable, exhibiting similar results to those shown in **Figure 3** for R = 4 . **Figure 5** presents the reconstructed images using reference and estimated ESPIRiT maps with acceleration factor R = 4 for another two subjects with large roll rotation and large overall translation, respectively. The reconstructed images for subject with the large roll rotation using estimated maps were highly consistent to those using reference maps at all slice locations visually and quantitatively (**Figure 5A**). No significant difference was observed. **Figure 5B** presents the reconstructed images with overall large head translation using reference and estimated maps, again showing similar performance at all slice locations.

**Figure 6** shows the reconstructed images using reference and estimated ESPIRiT maps with acceleration factor R = 4 for two subjects with relatively large and small head dimension along the left-right phase encoding direction. Despite the horizontal head dimension difference, the reconstructed images using estimated maps were in consistent agreement with those using reference maps at all slice locations. Similar noise amplification and artifact levels were observed. Note that, at high acceleration R = 4, the reconstruction images using reference ESPIRiT maps produced slightly less aliasing than that using estimated reference maps, but exhibited slightly more noise amplification. Together, these results demonstrated the robustness of our axial data trained deep learning model in reconstructing various axial datasets with different subject-coil geometries, even at very high acceleration.

The reconstruction results of the cardiac data are shown in **Figure 7**. The reconstruction performance using estimated ESPIRiT maps were observed to be comparable to that using reference maps at acceleration factor R = 2, 3 and 4. These results demonstrated that our method could adapt to the reconstruction of cardiac data, where there are often signal voids and relatively more rapid coil sensitivity variations when compared to brain data.

The proposed ESPIRiT reconstruction was also compared to SENSE reconstruction using coil sensitivity maps with and without the masking procedure (**Supporting Information Figure S2**). Consistent with previous results, reconstruction performance using reference and deep learning estimated ESPIRiT maps was comparable. With masking, SENSE reconstruction performance was worse when compared to ESPIRiT reconstruction using the reference or estimated map as expected because ESPIRiT reconstruction typically exhibits more robustness than the traditional SENSE reconstruction[8]. Without masking, SENSE reconstruction largely failed, underscoring the importance of masking operation for noise and artifacts suppression in SENSE reconstruction[50]. These results demonstrated our proposed deep learning calibrationless ESPIRiT reconstruction outperformed the traditional calibration-based SENSE reconstruction, if not similar.

# DISCUSSION

In this study, we present a deep learning framework to directly estimate multi-channel ESPIRiT maps from uniformly-undersampled multi-slice multi-channel MR data and then use such estimated ESPIRiT maps for subsequent image reconstruction. The proposed deep learning model is capable of predicting multi-channel ESPIRiT maps from uniformly-undersampled k-space data. The resulting maps are highly comparable to the reference ESPIRiT maps computed from 24 consecutive central k-space lines. Such deep learning estimated ESPIRiT maps led to excellent ESPIRiT reconstruction performance even at high acceleration, exhibiting a level of errors and artifacts consistently similar to that by using the reference ESPIRiT maps. Although coil sensitivity maps can be often acquired within ~10 seconds during pre-scan on most clinical scanners, our proposed method can be advantageous in some scenarios where accurate coil sensitivity maps are difficult to acquire, e.g., due to geometric mismatch between coil calibration data and image data as a result of motion or image distortion.

### Direct ESPIRiT map estimation by deep learning

Our proposed model can predict the ESPIRiT maps from uniformly-undersampled MR data through deep learning. It provides the ESPIRiT maps as prior knowledge for ESPIRiT reconstruction without any actual coil sensitivity calibration. This strategy differs from the majority of existing MRI deep learning models that aim to directly realize end-to-end reconstruction in image space[51], k-space[52-54] or cross-space[39,55].

Learning a direct nonlinear mapping from undersampled data to ESPIRiT maps is plausible and intuitively reasonable because ESPIRiT maps are smooth and only represent dominant information in coil sensitivity profiles[8]. As a result, they contain sparse information though they are 2D. ESPIRiT maps can be mostly represented linearly by calibration data, and compact convolutional kernels in deep learning architecture could effectively capture such linearity or even nonlinearity from undersampled data to form ESPIRiT maps[56,57].

The proposed deep learning model for estimating ESPIRiT maps are trained in a coil and geometry specific manner. In this study, we demonstrate its application to multi-slice axial brain datasets or multi-slice cardiac datasets separately acquired by the same MR receiving coil system, from which axial training datasets are acquired. Its estimation of ESPIRiT maps is robust even in presence of the unavoidable variation in subject-coil geometry when scanning different subjects in typical clinical MRI setting. For application to datasets acquired in other standard orientations, models should be trained accordingly with respective datasets.

### Existing deep learning method for coil sensitivity information estimation

DeepSENSE[33] is a recently proposed deep learning method that is designed to predict the coil sensitivity maps with high resolution from low-resolution coil sensitivity maps in a coil specific manner. Our method differs from DeepSENSE in two regards. First, our proposed model aims to estimate ESPIRiT maps. Coil sensitivity maps contain more information than ESPIRiT maps because only dominant information in coil sensitivity is preserved in ESPIRiT maps after the thresholding operation in SVD of coil calibration data[8]. Therefore, it is inherently easier and more reliable to estimate ESPIRiT maps as in this study than coil sensitivity maps by deep learning. Second, DeepSENSE still requires a limited number of consecutive central k-space lines for estimating low-resolution coil sensitivity maps as the input to the model. Our method does not need any such consecutive central k-space lines (or any calibration data associated with the dataset), thus providing a truly calibrationless approach to image reconstruction. Together with uniform undersampling, such complete elimination of calibration data requirement will benefit many routinely used clinical imaging protocols, such as FSE[58] and EPI[14,59], for acceleration or artifact correction.

### Need for incorporating subject-coil geometry information in deep learning

Our model does not utilize the subject-coil geometry information explicitly by designating it as a separate model input. Instead, we exploit it implicitly during the model training process through a hybrid loss function. Specifically, we first transform unaliased multi-slice image data to the standard reference multi-slice axial stack (i.e., a standard axial coordinate with fixed geometric relation to MR coil system), find the ESPIRiT maps at corresponding slice locations in the standard coordinate (i.e, $E^{ij}_{transformed}$), and use them as part of the hybrid loss during minimization process. To illustrate the effect of utilizing subject-coil geometry information in such manner, we have assessed the reconstruction performance using the models trained with and without such hybrid loss (**Supporting Information Figure S3A**). The results indicated that the reconstruction performance using model trained with hybrid loss is better than that by model trained without hybrid loss (i.e., $\lambda$ is fixed to 1). Additionally, reconstruction results using hybrid loss and single loss ($\lambda = 1$ and $\lambda = 0$) corresponding to Figure 4 is shown in **Supporting Information Figure 3B**. The reconstruction performance using ESPIRiT maps estimated from hybrid loss is better than that using ESPIRiT maps both estimated from single loss. The hybrid loss concept in **Figure 1B** and Equation (1) is designed to exploit the subject-coil geometric information readily accessible for each clinical MRI dataset.

## **Limitations and future directions**

Several limitations exist for our proposed deep learning reconstruction framework. First, our model training and application are coil specific. This may not pose a severe restriction as we envision that, in era of data-driven computing, truly effective MRI scanner should move towards self-learning, i.e., constant performance improvement through learning from the data generated by itself. Second, we incorporate the subject-coil geometry parameters into the model training process to improve the learning process. Imposing a hybrid loss in our model implicitly assumes that ESPIRiT maps in original and transformed locations (i.e., $E_{original}$ and $E_{transformed}$) are similar to certain extent due to geometric proximity of their locations. Therefore, both model and its application are orientation specific. In this study, we demonstrate the training and utility of our proposed model using typical multi-slice axial brain datasets and multi-slice oblique cardiac datasets. For other orientations, models should be trained accordingly with respective datasets acquired with the same multis-slice geometries and similar orientations. Third, we demonstrate our method here using T1-weighted MR brain data for both training and testing. We actually attempted to apply our model trained by T1-weighted MR brain data to reconstruct undersampled T2-weighted brain data, but with unsatisfactory performance. This indicates that our model is partially contrast or/and sequence specific. This limitation can be potentially mitigated by training the model with multi-contrast MR data[60-62]. **Supporting Information Figure S5** shows the preliminary results of training a single model for both T1- and T2-weighted image reconstruction, supporting such possibility. Fourth, we use multi-slice 2D MR data for estimating multi-slice 2D ESPIRiT maps. It will be desirable in the future to formulate deep learning strategy to predict 3D ESPIRiT maps directly so to tackle the 3D acquisitions scenarios where undersampling can occur in two dimensions. Opportunities can exist here due to (i) spatial smoothness nature of 3D ESPIRiT maps; and (ii) extra information redundancy along the slice-encoding direction since the neighboring slices have similar coil sensitivity information. Both could be exploited to facilitate model learning and achieve high acceleration likely at the expense of high computational demands. Fifth, model training is time-consuming though transfer learning can be explored in the future to alleviate the computational burden by improving model convergency[63]. Sixth, the performance of our proposed method remains to be further evaluated. For example, NRMSE and PSNR alone may not present comprehensive assessment of the image quality. Thus more experimentation and analysis are imperative. Last, our model needs the fully-sampled data to generate $E_{original}$ and $E_{transformed}$ for training. This requirement can be met by acquiring certain number of fully-sampled datasets first for each imaging protocol. However, this may not be convenient for some protocols such as single-shot EPI. One alternative to acquiring fully-sampled k-space datasets is to directly utilize the alias-free image datasets already reconstructed using existing parallel imaging methods.

## CONCLUSION

This study presents a framework to estimate ESPIRiT maps directly from uniformly-undersampled MR data using a deep learning model. The model is trained using fully-sampled datasets acquired with similar multi-slice geometry using the same MR receiving coil system. The proposed model can robustly predict the multi-channel ESPIRiT maps from uniformly-undersampled k-space data itself. They were highly comparable to the reference ESPIRiT maps computed from 24 consecutive central k-space lines. These deep learning estimated ESPIRiT maps lead to excellent ESPIRiT reconstruction performance even at high acceleration, exhibiting a similar level of errors and artifacts to that by using reference ESPIRiT maps. Our proposed framework offers a general strategy for calibrationless parallel imaging reconstruction through learning from coil and protocol specific data. It is highly applicabe to the application scenarios where accurate coil sensitivity calibration is difficult.

## ACKNOWLEDGEMENTS

This work was supported in part by Hong Kong Research Grant Council (R7003-19F, HKU17112120, HKU17127121 and HKU17127022 to E.X.W., and HKU17103819, HKU17104020 and HKU17127021 to A.T.L.L.), Lam Woo Foundation and Guangdong Key Technologies for Treatment of Brain Disorders (2018B030332001) to E.X.W.

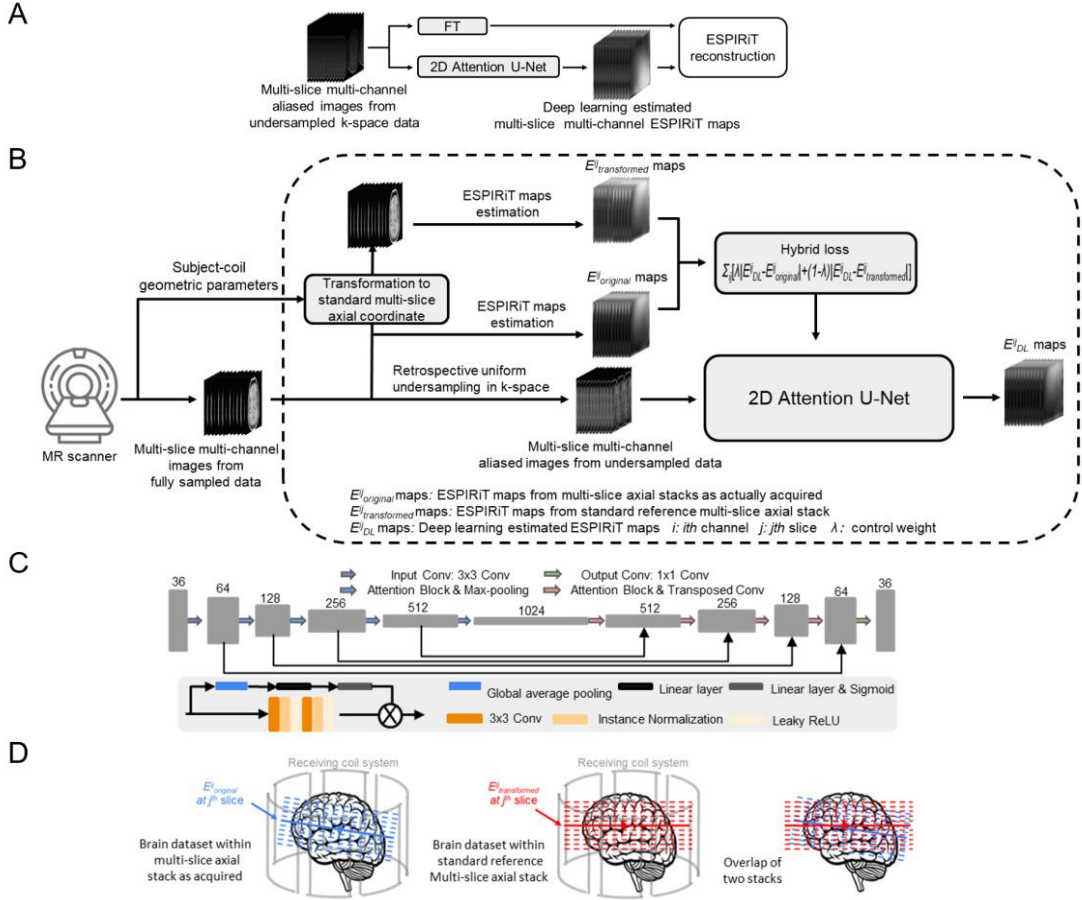

**Figure 1**. The proposed deep learning framework for ESPIRiT reconstruction without any coil sensitivity calibration. (**A**) ESPIRiT image reconstruction using deep learning estimated ESPIRiT maps. The trained model estimates the ESPIRiT maps directly from uniformly-undersampled k-space data, which are then used to perform L1-ESPIRiT reconstruction. (**B**) The proposed model and its training. The input is the multi-slice aliased image data reconstructed from undersampled axial multi-channel k-space data while the output is the multi-channel ESPIRiT maps. The training incorporates a hybrid loss on ESPIRiT maps at the original locations (i.e., $E^{ij}_{original}$) and their transformed locations (i.e., $E^{ij}_{transformed}$) within the standard reference multi-slice axial stack. (**C**) The architecture of the U-Net model. (**D**) Brain dataset within the original multi-slice axial stack as actually acquired (in blue where $E^{ij}_{original}$ maps reside on each slice) and the standard reference multi-slice axial stack (in red where $E^{ij}_{transformed}$ maps reside on each slice). Two stacks have identical slice gap/thickness geometry except their slightly different stack orientations and positions. For training, $E^{ij}_{transformed}$ maps for each brain dataset are obtained by first transforming the original multi-slice unaliased image stack to the standard reference multi-slice axial stack, and then computing the ESPIRiT maps at the corresponding slice locations. The transformation is performed through rigid-body rotation and translation using the subject-coil geometric parameters available in DICOM header. Note that the standard reference axial stack here has a fixed geometric relation to the MR coil system.

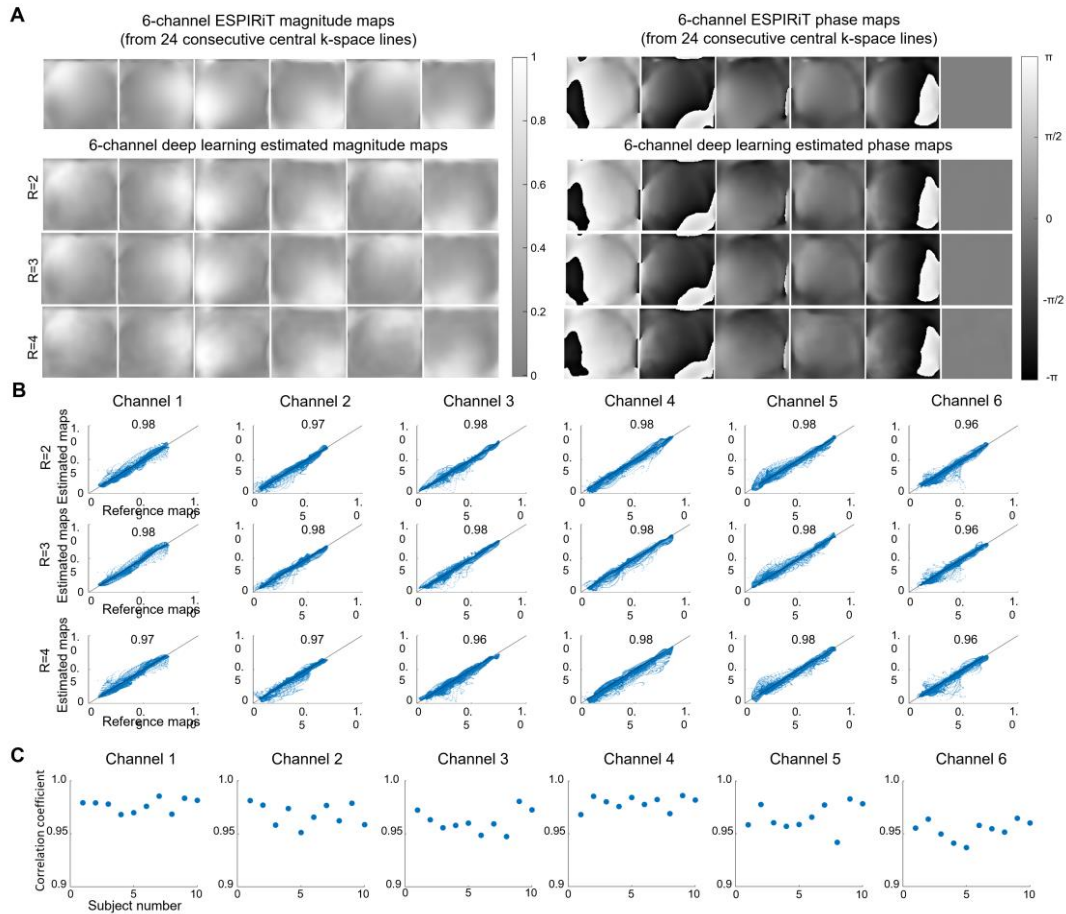

**Figure 2.** Reference ESPIRiT maps and deep learning estimated ESPIRiT maps (R = 2, 3 and 4 with number of channels = 6) for one subject. (**A**) Reference and deep learning estimated ESPIRiT magnitude and phase maps across 6 channels. The reference ESPIRiT maps are derived from 24 consecutive central k-space lines. (**B**) Pixel-wise magnitude correlation analysis of the reference vs. deep learning estimated ESPIRiT maps across 6 channels. Pearson correlation is used to quantify the similarity. (**C**) Correlation coefficients between deep learning estimated and reference ESPIRiT magnitude maps from one slice at similar locations among all 10 test subjects in 6 channels.

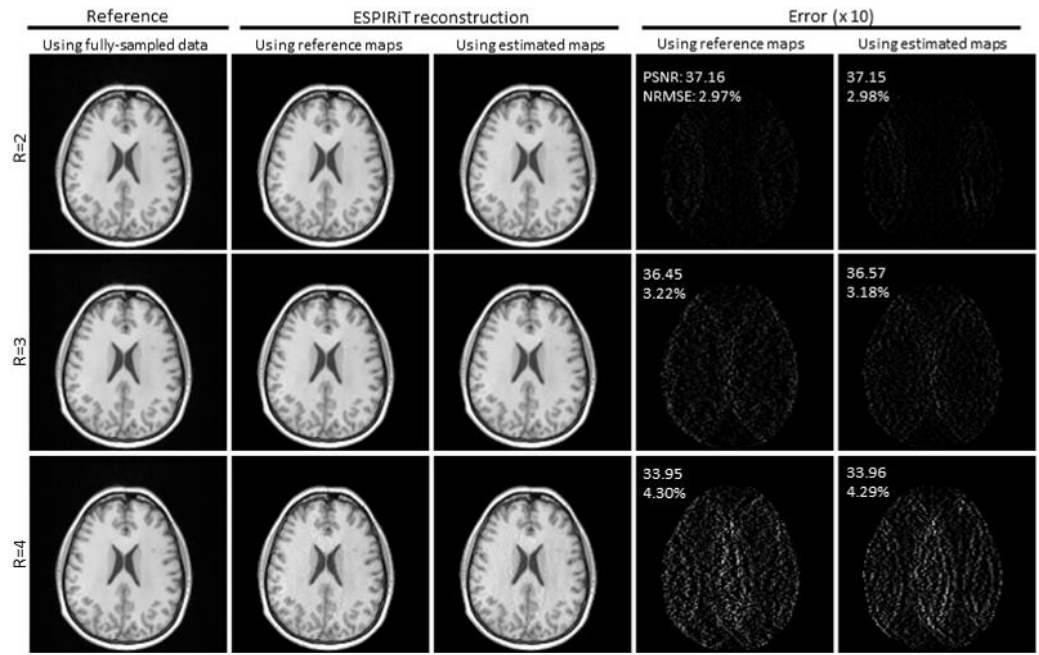

**Figure 3.** ESPIRiT reconstructed images with R = 2, 3 and 4 with number of channels = 6, corresponding to the reference and deep learning estimated maps shown in Figure 2. Error images, i.e., difference between reconstructed images and reference images are also shown together with the PSNR and NRMSE values. PSNR: peak signal to noise ratio. NRMSE: normalized root mean squared error.

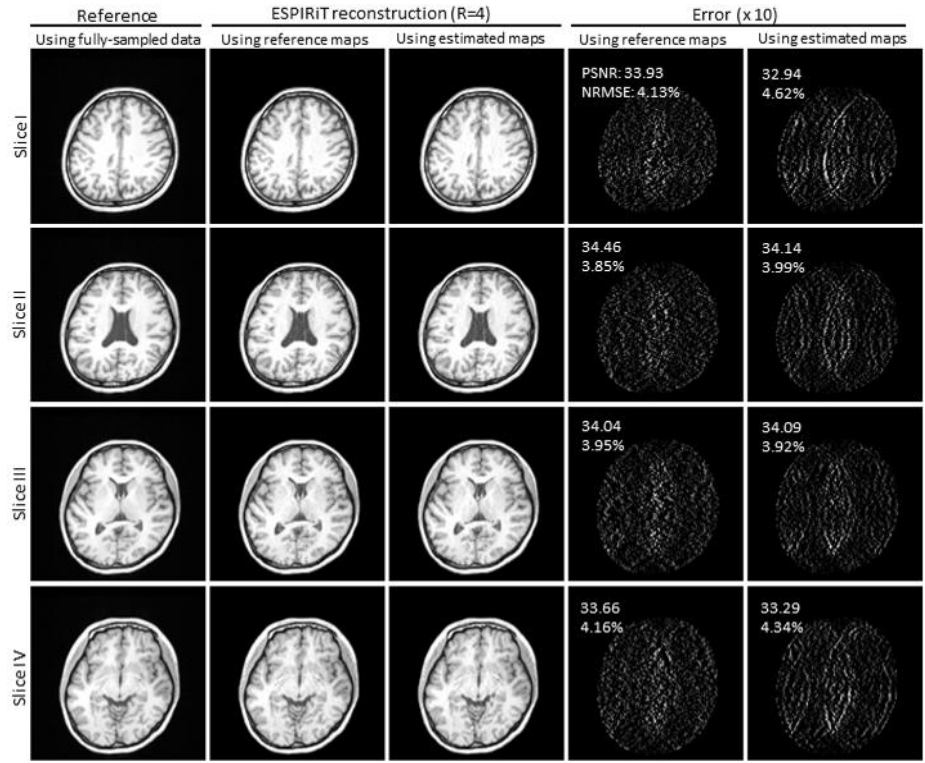

**Figure 4.** Reconstruction performance for one subject with a large pitch rotation (about -10 degree) using reference and deep learning estimated ESPIRIT maps at R = 4 with number of channels = 6.

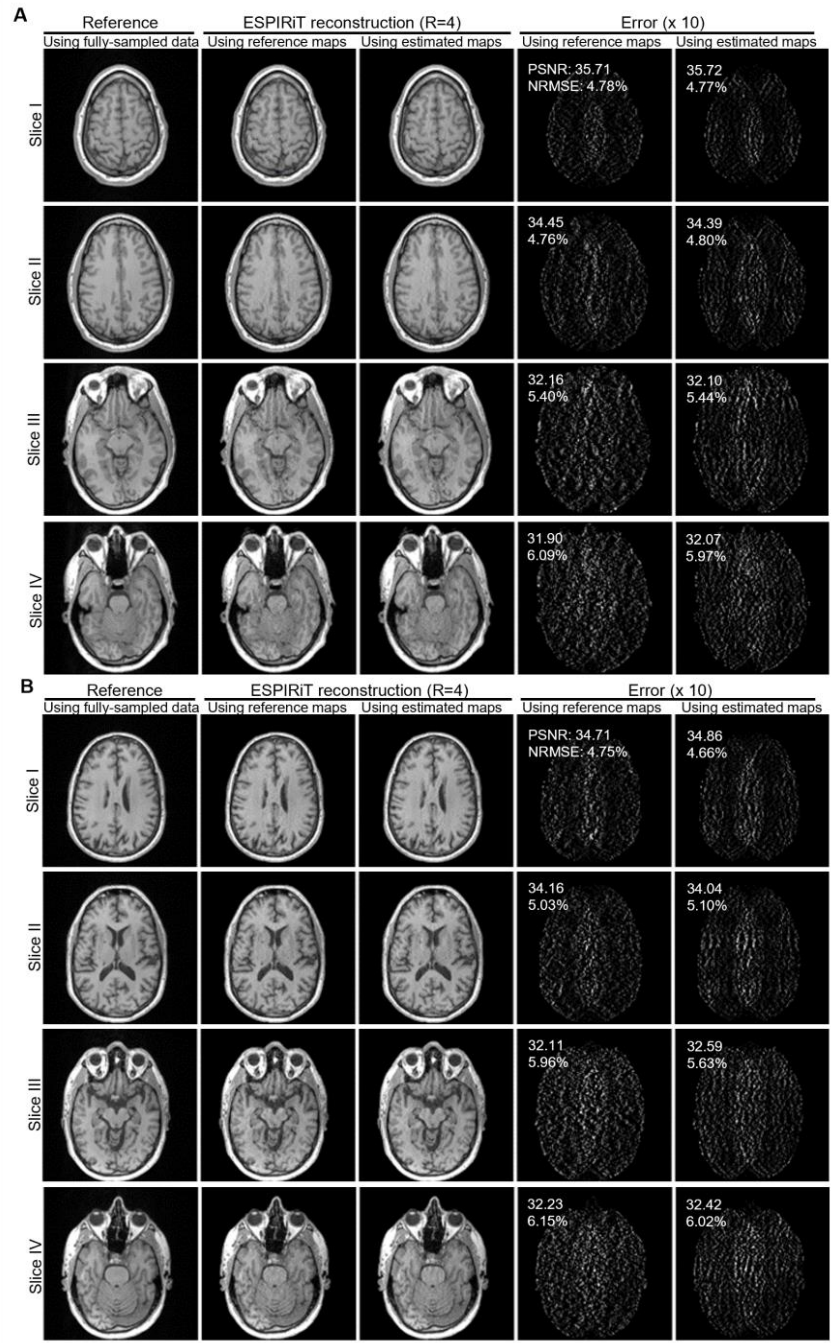

**Figure 5.** Reconstruction performance using reference and deep learning estimated ESPIRIT maps (R = 4 with number of channels = 6) for two subjects with large roll rotation and large overall translation, respectively. (**A**) Reconstructed images with the large roll rotation (+ 6 degrees) (**B**) Reconstructed images with the large overall head translation with summation of translations along three dimensions being 18 pixels.

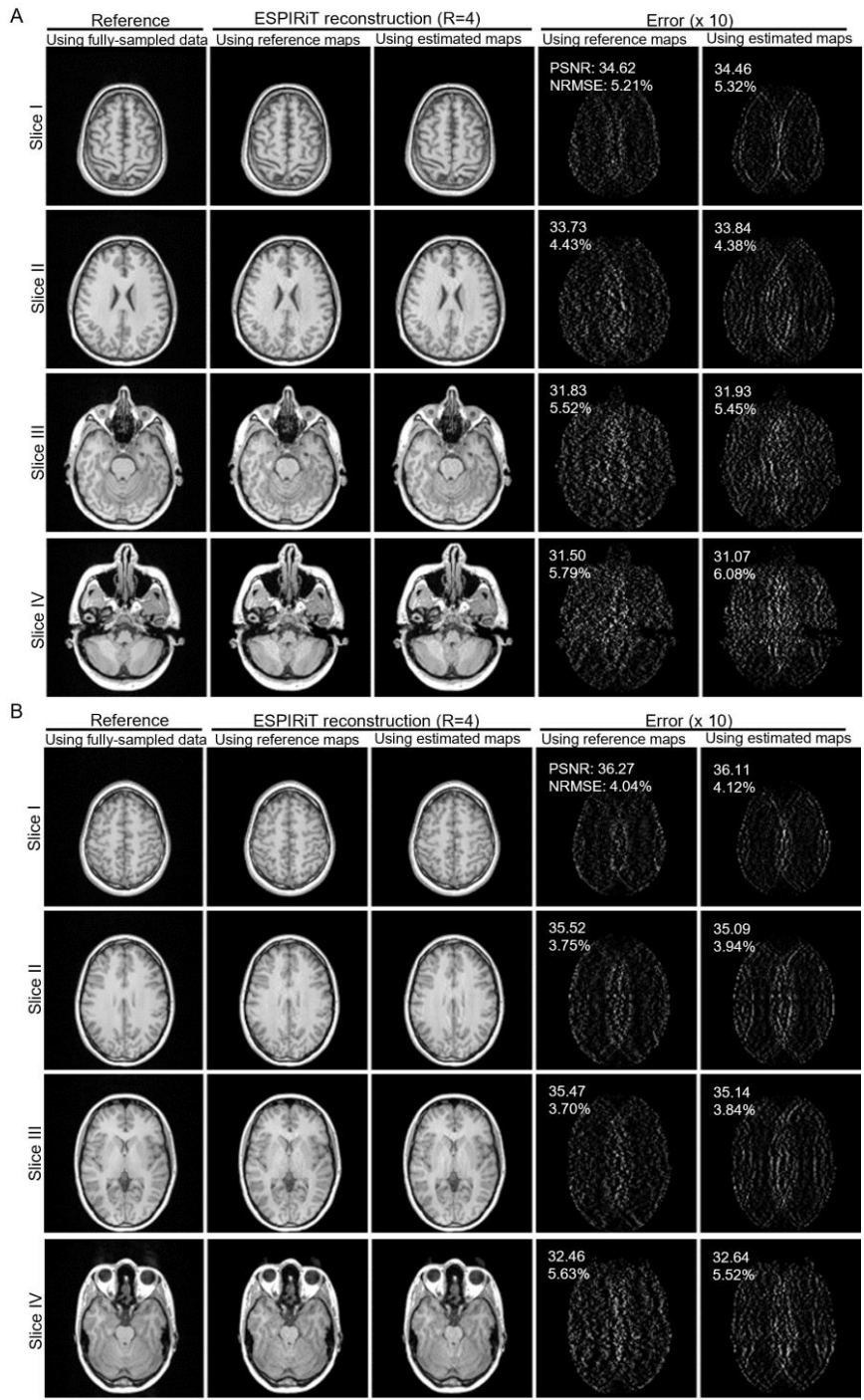

**Figure 6.** Reconstruction performance for two subjects with large and small head dimension along left-right phase-encoding direction using reference and deep learning estimated ESPIRiT maps at R = 4 with number of channels = 6. (**A**) The reconstructed images for one subject with a large left-right head dimension. (**B**) The reconstructed images for one subject with a small left-right head dimension.

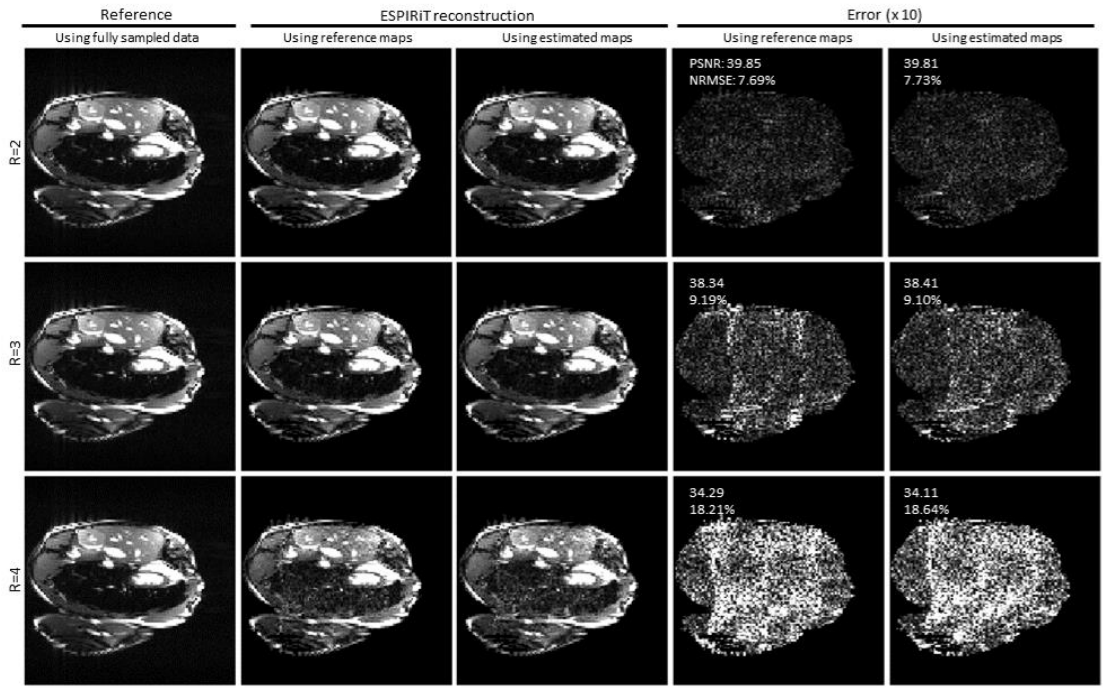

**Figure 7.** Reconstruction performance of the cardiac data from one subject using ESPIRiT reference maps and deep learning estimated maps with R = 2, 3 and 4 with channel number =  6.

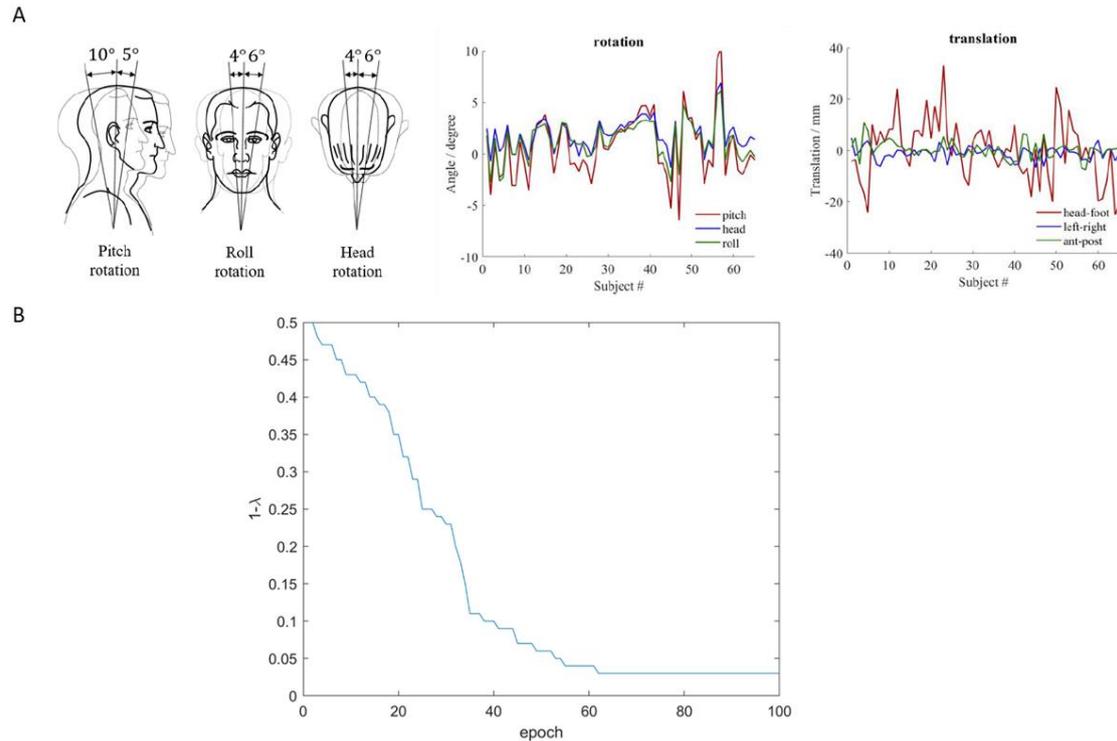

**Supporting Information Figure S1.** (**A**) Summary of the subject-coil geometric parameters in terms of 3 rotations and 3 translations among all 65 subjects whose datasets were used in this study. The rotation ranges/distributions and translation distribution are displayed. These subject-coil geometry parameters were obtained from the DICOM tags (0020,0037) and (0020,0032), respectively, of each dataset. (**B**) The weight change of the 2$^{nd}$ term in Equation (1) during a typical model training for T1-weighted brain data. With each epoch, the model training process focuses more on the 1$^{st}$ term instead of the 2$^{nd}$ term, especially when the training finally converges.

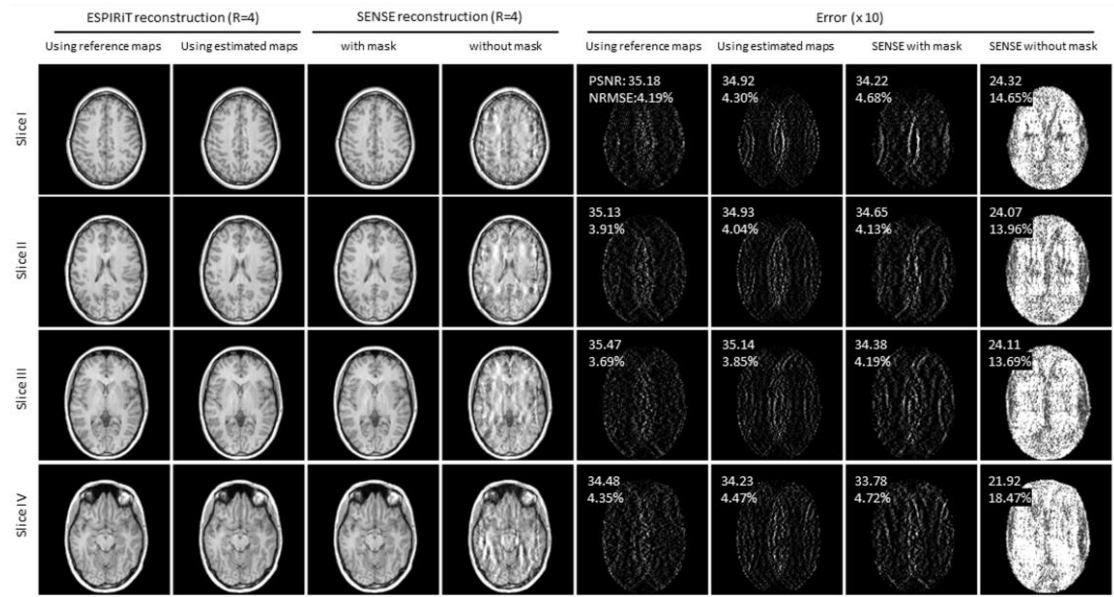

**Supporting Information Figure S2.** SENSE reconstruction with and without masking and ESPIRiT reconstruction using reference and deep learning estimated maps for one subject. Coil sensitivity maps are calculated from fully-sampled data using BART Toolbox in MATLAB. Masking procedure: the coil sensitivity maps are thresholded by pixel intensity and set to 0 if below the threshold in order to reduce noise propagation during SESEN reconstruction.

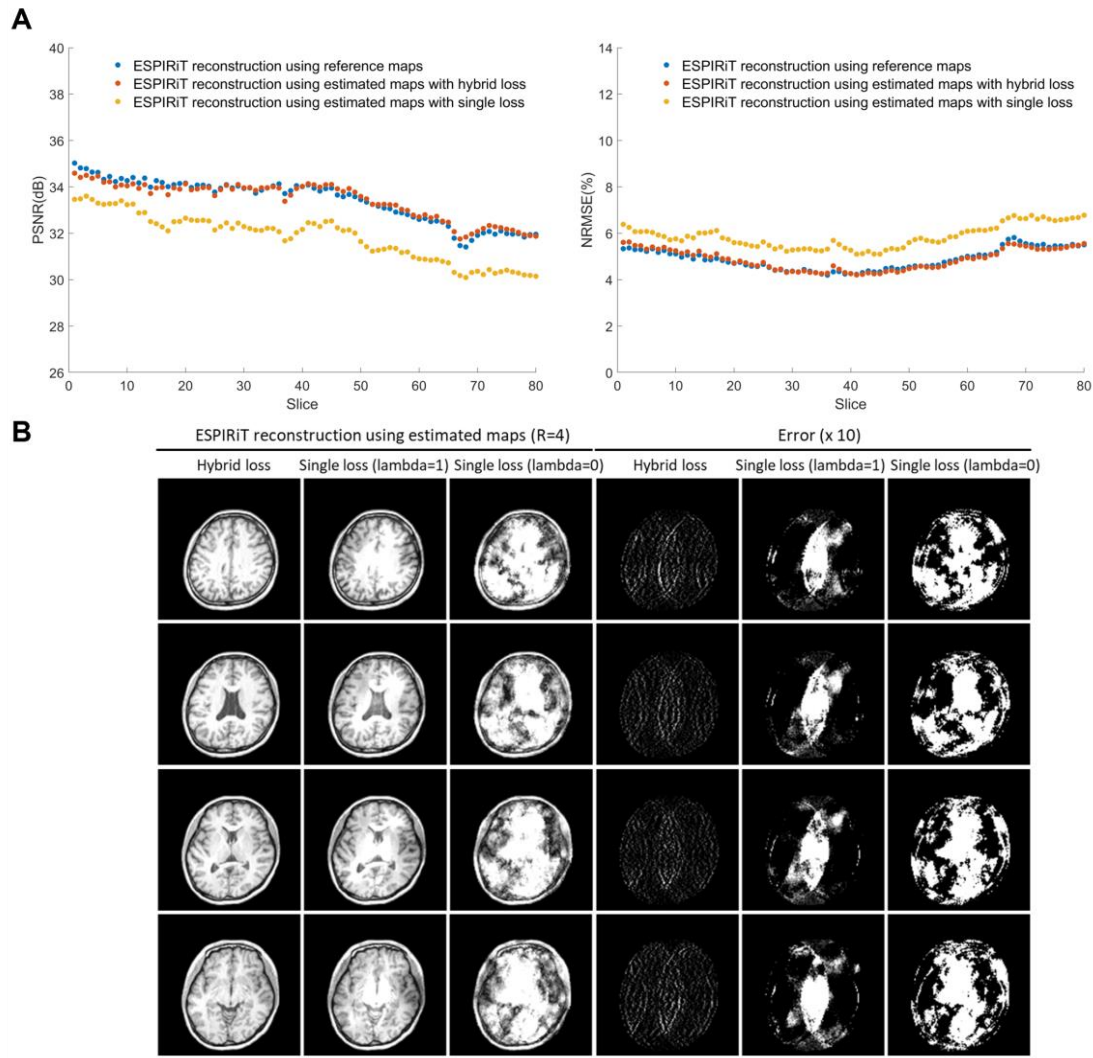

**Supporting Information Figure S3.** (**A**) Reconstruction performance evaluation with R = 4 and number of channels = 6 for two subjects corresponding to Figure 3 (**A**) and Figure 4 (**B**), respectively. The PSNR and NRMSE values are compared at all 80 slice locations for images reconstructed using the reference ESPIRiT maps, deep learning estimated ESPIRiT maps with the proposed hybrid loss, and deep learning estimated ESPIRiT maps with the simple loss (i.e., λ = 1). (**B**) Reconstruction performance of one subject using deep learning estimated ESPIRIT maps from hybrid loss, single loss (λ = 1) and single loss (λ = 0) at R = 4 with number of channels = 6.

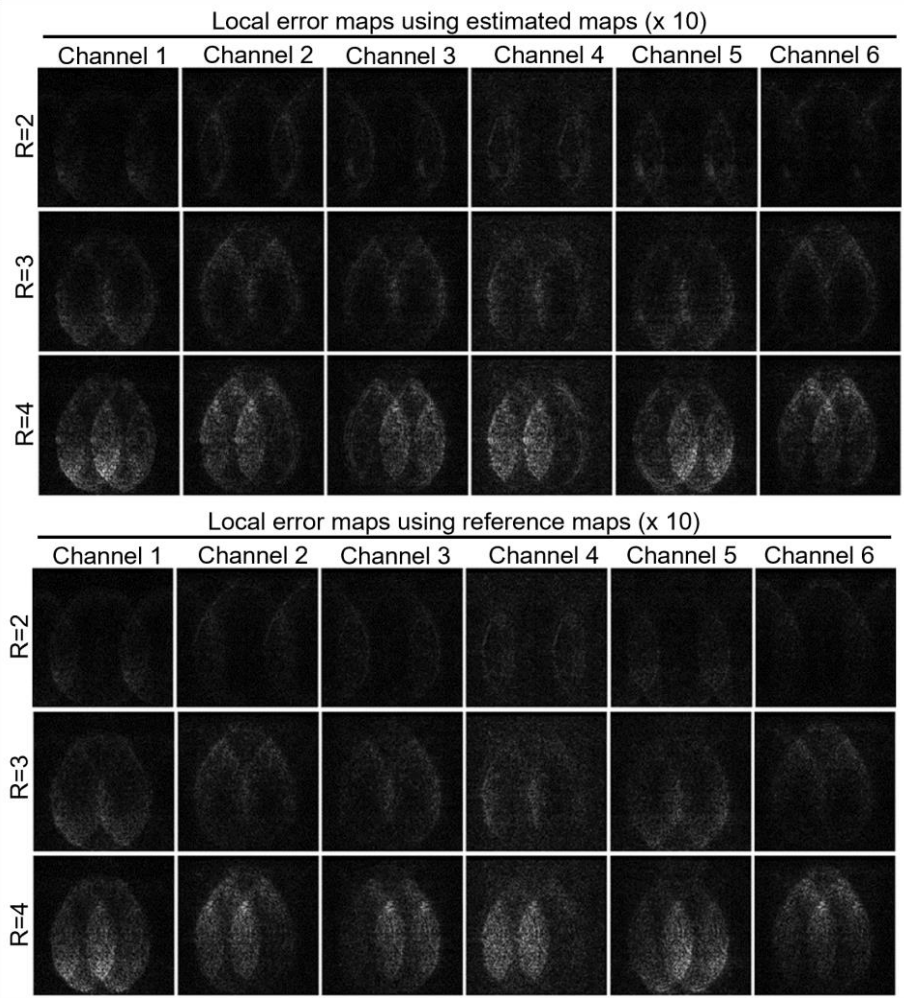

**Supporting Information Figure S4.** Local error maps of reconstruction results using reference and estimated maps with R = 2, 3 and 4 with the number of channels = 6 corresponding to Figure 3.

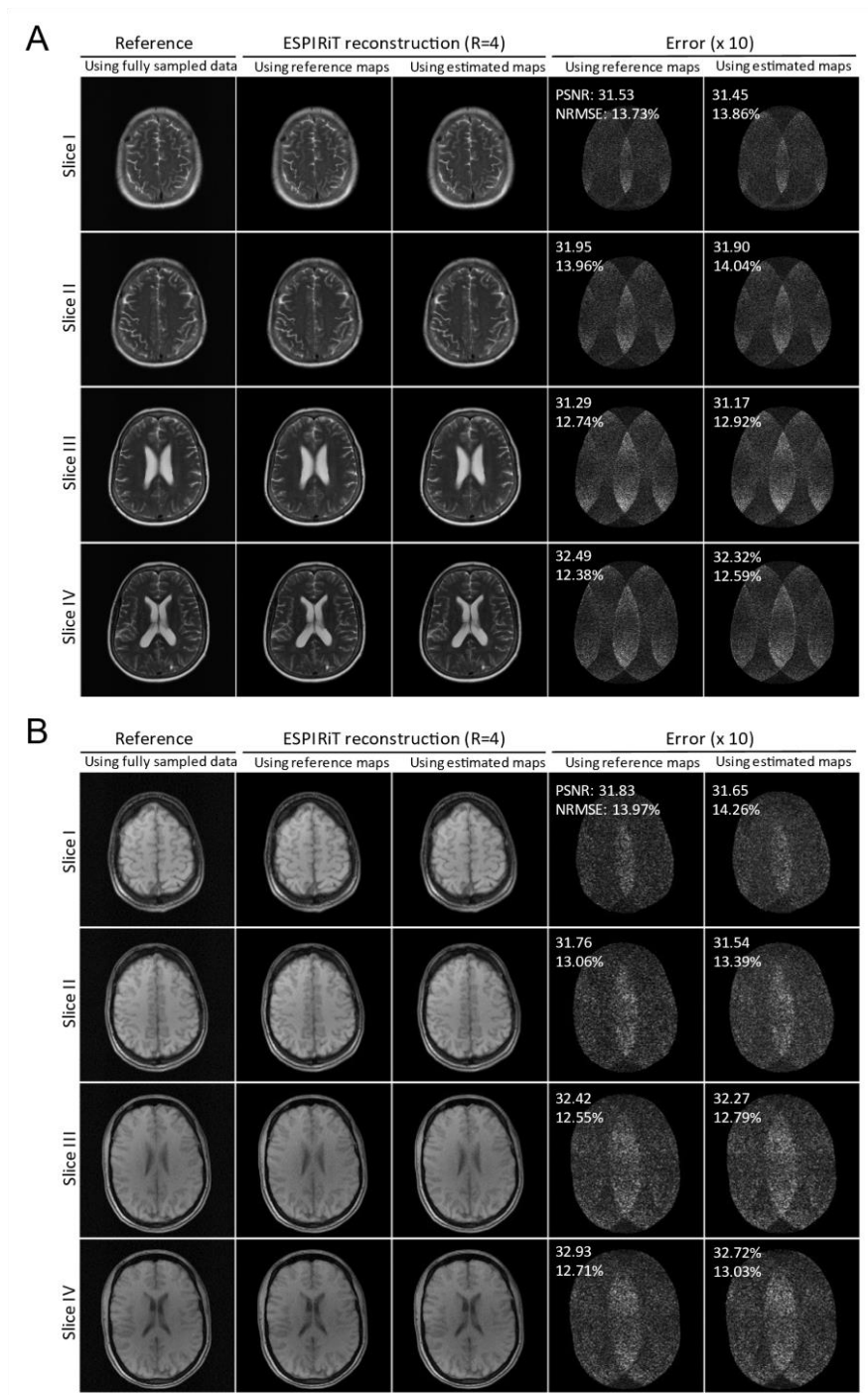

**Supporting Information Figure S5.** Preliminary results of reconstructing T1- and T2-weighed brain data at R = 4 using estimated ESPIRiT maps by deep learning. Here a single model was trained by both T2- and T1-weighted brain data, and applied to T1- or T2-weighted image reconstruction. The multi-slice 16-channel T1- and T2-weighed data (each from 40 different subjects) from fastMR database[64] were employed with model training time ~. Coil compression was used to compress the data to 6 channels. The matrix size is 250x250.